\documentclass[showpacs,showkeys,prc,nofootinbib]{revtex4}
\usepackage{exscale}
\usepackage{amsmath}
\usepackage{amssymb}
\usepackage[dvips]{graphicx}
\usepackage{float}
\newcommand{\bea}{\begin{eqnarray}}
\newcommand{\eea}{\end{eqnarray}}
\newcommand{\nn}{\nonumber}
\newcommand{\nnl}{\nonumber\\}
\begin{document}


\title{Spectral Function of Quarks in Quark Matter}
\thanks{Supported by DFG and GSI.}
\author{F. Fr\"omel}
\author{S. Leupold}
\author{U. Mosel}
\affiliation{Institut f\"ur Theoretische Physik, Universit\"at Giessen,
Heinrich-Buff-Ring 16, D-35392 Giessen, Germany}


\begin{abstract}
We investigate the spectral function of light quarks in infinite quark matter
using a simple, albeit self-consistent model. The interactions between the
quarks are described by the SU(2) Nambu--Jona-Lasinio model. Currently mean-field 
effects are neglected and all calculations are performed in the chirally
restored phase at zero temperature. Relations between correlation functions and
collision rates are used to calculate the spectral function in an iterative
process.
\end{abstract}

\maketitle


\section{Introduction}

It is well known that short-range correlations have influence on the properties
of nuclear matter and finite nuclei. The spectral functions observed in
$A(e,e'p)X$ \cite{x1} and $A(e,e'pp)X$ \cite{x2} experiments are much wider
spread than explained by mean-field dynamics. Due to short-range correlations
a substantial amount of high-momentum processes is contained in the spectral
function.

There have been many theoretical approaches, mainly based on nuclear many-body
theory, trying to understand the short-range correlations in nuclei. The
results of these calculations have converged in the last years \cite{ben,dick}.
In particular, the significant population of high-momentum states in the nucleon
momentum distribution which is an overall measure for short-range correlations
is described rather well.

Our present study is motivated by the work of Lehr et al. \cite{le}. They have
calculated the nucleon spectral function in nuclear matter in a simple model
based on transport theory. Direct relations between collision rates and
correlation functions form a self-consistency problem and were used to
determine the spectral function iteratively. In their calculations the drastic
assumption of an averaged, constant scattering amplitude was made. Hence the
results are dominated by the properties of the available phase space.

It has turned out that the results of Lehr et al. agree surprisingly well with
experimental data \cite{cda} and sophisticated "state-of-the-art" calculations
from many-body theory \cite{ben}. Spectral functions, momentum distributions,
occupation probabilities, and response functions are all in close agreement with
other calculations. It is in particular striking that the only parameter in
that model, the coupling strength, is enough  to describe size and slope of the
high-momentum tail in the nucleon momentum distribution in nuclear matter.

This leads to two conclusions. First, the properties of the nucleon spectral
function in nuclear matter are determined by phase space effects and an average
strength of the short-range correlations. Second, the detailed structure of the
interaction seems to be unimportant as long as the calculations are made in a
self-consistent framework.

Due to the success of this model, we take up the concept and make the assumption
that similar conditions apply for the short-range correlations in quark matter.
Only a few changes to the existing nucleon model are necessary to get a working
model for quark matter. First of all, the low current masses of up and down
quarks make it necessary to perform the calculations relativistically. This
implies some complications since the relativistic, fermionic spectral function
is a $4\times4$ matrix in spinor space. Using symmetry arguments it can be
shown that a reduction to three scalar functions is possible. This makes all
expressions look more complicated and increases the numerical efforts. However,
it does not change the fundamental concept of the model in any way.

In addition a sensible model for the quark interactions has to be used. Our
basic assumption is that the details of the interaction should be relatively
unimportant as long as the overall strength is correct and the relevant
symmetries are respected. Hence we use the well known Nambu--Jona-Lasinio model
(NJL model) in its SU(2) version in our calculations. It has the same
symmetries as QCD and describes an effective pointlike interaction with a
constant coupling strength. Usually the NJL model and its extensions are used
in a mean-field approximation. In the spirit of our works on nucleons, we
present here an approach that goes beyond the mean-field approximation in a
self-consistent way.

In the present work, we use the standard NJL model and do not consider an
additional attractive interaction in the quark-quark channel as it appears in
extended NJL models that are used in recent works on color superconductivity
\cite{csc}. The aspect of color superconductivity -- albeit interesting -- is
beyond the scope of the present work. A more detailed discussion can be found
in sec. \ref{sec:model} of this paper.

The calculations we present in the following are restricted to the chirally
restored phase where the quarks are massless. The relativistic structure of the
spectral function gets simpler in this phase, thus simplifying the analytic
expressions and reducing the time needed for the numerical calculations
significantly. We will come back to that point below.

In sec. \ref{sec:model} the theoretical aspects of the model are summarized.
It is shown how  expressions for the self-energy and the spectral function are
constructed that are directly related to each other. The resulting
self-consistency problem can be solved in an iterative approach. The technical
details and the results of our numerical calculations are presented in
sec. \ref{sec:results}. We show the results for the spectral function, its
width and the momentum distribution of the quarks. The role of short-range
correlations is investigated. Section \ref{sec:summary} summarizes our results
and gives an outlook.

\section{\label{sec:model}The Model}

\subsection{Basic relations}

In Ref. \cite{le}, the model we adopt is described for nonrelativistic nucleons.
Many of the details concerning the definition and meaning of the basic
quantities can also be found in Refs. \cite{kb,dan}. Some aspects of working with
relativistic nucleons are discussed in Ref. \cite{bo}. The fundamental elements of
our model are the one-particle Green's functions without ordering, $g^>$ and
$g^<$. For relativistic fermions they are defined as
\bea
	i g^>_{\alpha\beta}(1,1')  &=& \langle\psi_\alpha(1) \bar\psi_\beta(1')
	\rangle \nnl
	-i g^<_{\alpha\beta}(1,1') &=& \langle \bar\psi_\beta(1') \psi_\alpha(1)
	\rangle ,
\eea
where the $\psi$ are field operators in the Heisenberg picture. The indices
denote the spin degrees of freedom; isospin and color indices have been
supressed here. $1$ and $1'$ denote two space-time coordinates. Since we
describe fermions with our model the field operators are spinors and the
Green's functions are $4 \times 4$ matrices in spinor space.

At this point we also introduce the retarded Green's function $g^\mathrm{ret}$.
It is closely related to $g^\gtrless$,
\bea
	g^\mathrm{ret}_{\alpha\beta}(1,1')=\Theta(t_{1}-t_{1'})
	\left[g^>_{\alpha\beta}(1,1')-g^<_{\alpha\beta}(1,1') \right].
\eea
The Fourier-transformed Green's functions are used to define the spectral
function $\mathcal{A}(p)$. Obviously $\mathcal{A}(p)$ takes over the matrix
structure of the Green's functions
\bea
	\mathcal{A}_{\alpha\beta}(p)
	 =i[g^>_{\alpha\beta}(p)-g^<_{\alpha\beta}(p)]
	 =-2\mathrm{Im}g^\mathrm{ret}_{\alpha\beta}(p),
	\label{eq:spec-def}
\eea
where "Im" is given by
$\mathrm{Im}F=\frac{1}{2i}(F-\gamma_0 F^\dagger\gamma_0)$,
and correspondingly
$\mathrm{Re}F=\frac{1}{2}(F+\gamma_0 F^\dagger\gamma_0)$.

In addition to eq. (\ref{eq:spec-def})
the Green's functions are related to the spectral function via the
energy-momentum space distribution function. In thermal equilibrium one has
\bea
-ig^<_{\alpha\beta}(p)&=&\mathcal{A}_{\alpha\beta}(p)n_F(p_0),\label{eq:gkl-spec}
\\
 ig^>_{\alpha\beta}(p)&=&\mathcal{A}_{\alpha\beta}(p)[1-n_F(p_0)],
 \label{eq:ggr-spec}
\eea
with the thermal Fermi distribution $n_F(p_0)$. At zero temperature the Fermi
distribution becomes a step function, $n_F(p_0)=\Theta(\omega_F-p_0)$.

The single particle self-energy of an interacting quantum system can be split
into a mean-field part $\Sigma^\mathrm{mf}$ and the collisional self-energies
$\Sigma^\gtrless$. Like the spectral function the self-energy has a
matrix structure in spinor space:
\bea
	\Sigma_{\alpha\beta}(1,1')=	\Sigma^\mathrm{mf}_{\alpha\beta}(1,1')
	+\Theta(t_{1}-t_{1'})\Sigma^>_{\alpha\beta}(1,1')+\Theta(t_{1'}-t_{1})
	\Sigma^<_{\alpha\beta}(1,1').
\eea
The mean-field self-energy is time local,
$\Sigma^\mathrm{mf}\sim\delta(t_{1}-t_{1'})$. It describes the effects
corresponding to the motion of noninteracting particles in a mean-field
potential and is responsible for the dynamical generation of the constituent
quark masses. The collisional part of the self-energy contains the effects that
arise from decays and particle collisions in the medium.

As for the Green's functions a retarded self-energy is introduced
\bea
	\Sigma^\mathrm{ret}_{\alpha\beta}(1,1')=
	\Sigma^\mathrm{mf}_{\alpha\beta}(1,1')+\Theta(t_{1}-t_{1'})
	\left[\Sigma^>_{\alpha\beta}(1,1')-\Sigma^<_{\alpha\beta}(1,1')\right].
\eea
In momentum space the width of the spectral function is given by the imaginary
part of this retarded self-energy. One has
\bea
	\Gamma_{\alpha\beta}(p)=-2\mathrm{Im}\Sigma^\mathrm{ret}_{\alpha\beta}(p)
	=i[\Sigma^>_{\alpha\beta}(p)-\Sigma^<_{\alpha\beta}(p)].
	\label{eq:gamma-sigma}
\eea
$\Sigma^\mathrm{mf}$ does not contribute to the imaginary part of
$\Sigma^\mathrm{ret}(p)$ since it is time local. Hence the width of the
spectral function is completely given by the collisional self-energies
$\Sigma^\gtrless$.  In \cite{kb} it has been shown that $i\Sigma^>(p)$ and
$-i\Sigma^<(p)$ are identical to the total collision rates for scattering into
(gain rate) and out of (loss rate) the configuration
$(p_0,\vec p)$, respectively.

From analyticity it follows that the real part of the self-energy is related
to $\Gamma$ by a dispersion relation
\bea
	\mathrm{Re}\Sigma^\mathrm{ret}(p)=\Sigma^\mathrm{mf}(p)
	+\mathcal{P}\int\frac{dp_0'}{2\pi}\frac{\Gamma(p_0',\vec p)}{p_0-p_0'}.
	\label{eq:disp-rel}
\eea
If the width is known over the full energy range the real part of
$\Sigma^\mathrm{ret}$ can be calculated dispersively.

\subsection{Interactions in the NJL Model}

In the following, we use the standard SU(2) version of the
NJL model \cite{klev,klvw}. It has been designed to resemble the symmetries
of QCD. In particular, it allows the dynamical generation of fermion masses via
the mechanism of chiral symmetry breaking. The Lagrangian of this effective
interaction is given by
\bea
   \mathcal{L}_\mathrm{NJL}= \bar \psi i \partial\!\!\!/ \psi
   + G [(\bar \psi \psi)^2+(\bar \psi  i \gamma_5 \vec \tau \psi)^2],
   \label{eq:njl-lagr}
\eea
where $G$ is the coupling strength, independent of energy and momentum, and
the $\tau_i$ are the isospin Pauli matrices. The small current quark masses
are neglected here.

The pointlike interaction renders this model nonrenormalizable. In our
calculations we deal with this problem by introducing a cutoff for the
three-momenta. This cutoff and the coupling constant are the free parameters
of the NJL model. Usually their values are chosen so that the model reproduces
the correct values for the quark condensate and the pion decay constant in
vacuum.

Before we continue with the discussion of our present model we want to comment
on its relation to other present investigations. Recent works on color
superconductivity \cite{csc,berg,lang} consider an extended NJL model.
In the case of two flavors the interaction is given by
\bea
        \mathcal{L}_\mathrm{int}=G_1 [(\bar \psi \psi)^2+(\bar \psi
        i \gamma_5 \vec \tau \psi)^2]
        +G_2 (\bar \psi i \gamma_5 \tau_2 \lambda_2 \bar \psi^T)
        (\psi^T C i \gamma_5 \tau_2 \lambda_2 \psi), \label{eq:ext-njl}
\eea
where $C$ is the charge conjugation operator and $\lambda_2$ is a Gell-Mann
matrix in color space. $G_1$ and $G_2$ are coupling constants. In the 
mean-field approximation, the new term on the right of eq. (\ref{eq:ext-njl}) 
introduces an attractive interaction in the quark-quark channel that can lead 
to a pairing of quarks at the Fermi surface. This pairing breaks the color 
symmetry spontaneously and gives rise to color superconductivity. We note in 
the passing that the use of such an extended Lagrangian is not free 
from ambiguities \cite{berg,lang} due to the freedom of Fierz rearrangements.

As we will discuss below, we go beyond the mean-field approximation in our
approach. We have decided as a first step to construct a consistent
model based on the well established Lagrangian (\ref{eq:njl-lagr}). An
additional attractive quark-quark channel is not considered at this stage.
Color superconductivity is definitely an interesting phenomenon and can be
incorporated into the model in the future.

\subsection{Structure of the relativistic expressions}

As shown above the spectral function of relativistic fermions has a matrix
structure. The most general form of this structure is found by writing the
spectral function in terms of the 16 linear independent products of the
$\gamma$ matrices (Clifford algebra) and demanding the invariance under certain
symmetries \cite{bd}. We consider here parity and time-reversal symmetry. A
strongly interacting system of infinite size and in thermal equilibrium is
invariant under these transformations. In the rest frame of the medium this
leads to the following form for the spectral function:
\bea
	\mathcal{A}(p)=\rho_\mathrm{s}(p_0,\vec p^{\,2})
	+\rho_\mathrm{0}(p_0,\vec p^{\,2})\cdot\gamma^0
	+\rho_\mathrm{v}(p_0,\vec p^{\,2})\hat{\vec p} \cdot \vec \gamma,
	\label{eq:spec-struc}
\eea
where $\hat{\vec p}$ is a unit vector in the momentum direction. As a first
step we restrict our calculations in the present work to the chirally restored
phase. Since the "scalar" term $\rho_\mathrm{s}$ is not chirally invariant it
has to be zero; the spectral function is then completely determined by the two
scalar functions $\rho_0$ and $\rho_\mathrm{v}$. Note that the spectral
function for free quarks with $p_0>0$  is given by
$\mathcal{A}=\pi |\vec p|^{-1}\! \not\!p \, \delta(p_0-|\vec p|)$. According to
our choice of signs in eq. (\ref{eq:spec-struc}) this means that $\rho_0>0$ and
$\rho_\mathrm{v}<0$ in this case.

The structure of the self-energies $\Sigma^\gtrless$ and the width $\Gamma$ are
identical to that of the spectral function. They too consist of three
scalar functions,
\bea
	\Sigma^\gtrless(p)&=&\Sigma^\gtrless_\mathrm{s}(p_0,\vec p^{\,2})
	+\Sigma^\gtrless_\mathrm{0}(p_0,\vec p^{\,2})\cdot\gamma^0
	+\Sigma^\gtrless_\mathrm{v}(p_0,\vec p^{\,2})\hat{\vec p} \cdot \vec \gamma,
\\
	\Gamma(p)&=&\Gamma_\mathrm{s}(p_0,\vec p^{\,2})
	+\Gamma_\mathrm{0}(p_0,\vec p^{\,2})\cdot\gamma^0
	+\Gamma_\mathrm{v}(p_0,\vec p^{\,2})\hat{\vec p} \cdot \vec \gamma,
\eea
with $\Sigma^\gtrless_\mathrm{s}=\Gamma_\mathrm{s}=0$ in the chirally symmetric
phase.

\subsection{Self-consistent expressions}

The mean-field part $\Sigma^\mathrm{mf}$ of the self-energy is given by the
tadpole diagrams. Currently chiral symmetry is restored "by hand" in our model.
This means that the system under consideration is forced into the chirally
restored phase by setting the constituent quark masses, i.e., the Hartree term
of $\Sigma^\mathrm{mf}$ to zero. The Fock term contains a part that needs not 
to be zero but this part can be absorbed in a redefinition of the chemical 
potential \cite{klev}. Note that
one can do this for any temperature and chemical potential since the gap
equation for the quark masses \cite{klev,klvw} is always solved for $m=0$.
However, thermodynamically this might not be the favored phase when there
exists also a finite solution for the gap equation.

\begin{figure}[htbp]
	\begin{center}
		\includegraphics{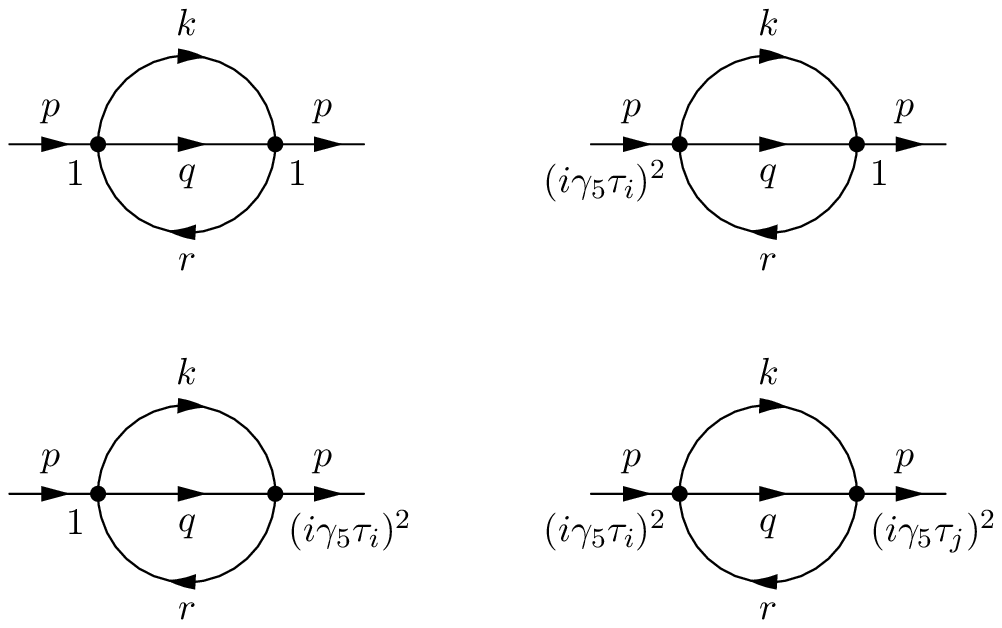}
	\end{center}
	\caption{\label{fig:selfenergy}Lowest order contributions to the self-energy
	$i\Sigma^>(p)$ in the SU(2) NJL model (Born diagrams). Since we have a
	pointlike interaction the direct and the exchange diagrams look the same in
	this figure.}
\end{figure}

\begin{figure}[htbp]
	\begin{center}
		\includegraphics{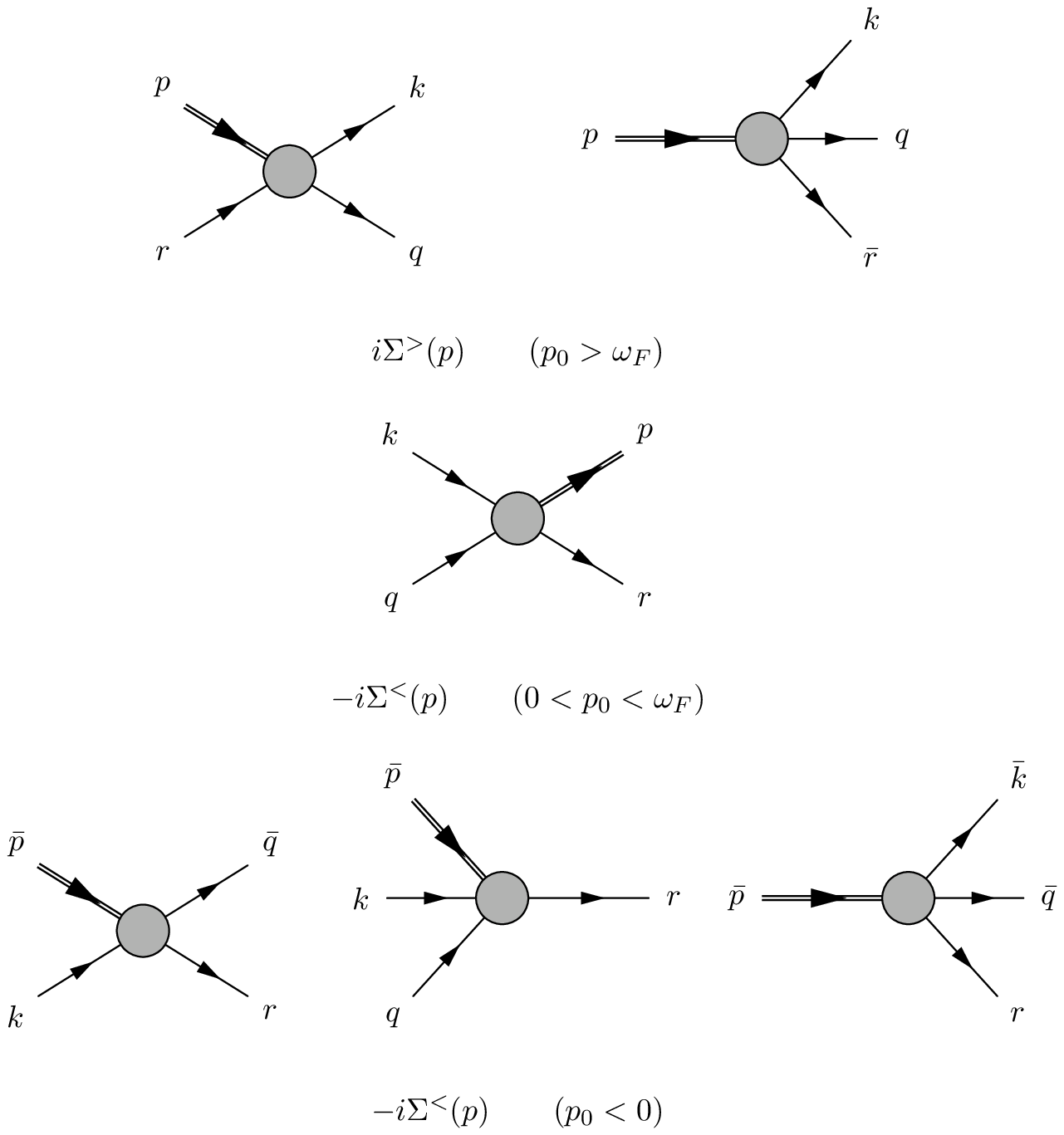}
	\end{center}
	\caption{\label{fig:scatter}Scattering processes corresponding to the
	self-energies $\Sigma^\gtrless(p)$ in the Born
	approximation. The double lines carry the four-momentum $p$ of
	$\Sigma^\gtrless$;
$p,q,k,r$ denote quarks with four-momentum $(p_0,\vec p)$, etc. (with
$p_0,q_0,k_0,r_0>0$), and $\bar p, \bar q,\bar k,\bar r$ denote antiquarks
with four-momentum $(-p_0,\vec p)$, etc. (with $p_0,q_0,k_0,r_0<0$).
Total collision rates are found by integrating over the four-momenta $k,q,r$
(respectively $\bar k,\bar q,\bar r$), i.e., all lines except the double line.
In the first row the scattering-out processes contributing to the loss rate
$i\Sigma^>$ are shown. The second row shows the only contribution to the gain
rate $-i\Sigma^<$ for positive $p_0$. In the last row the processes
contributing to the antiparticle loss rate $-i\Sigma^<$ at $p_0<0$ are shown.
Note that there are no contributions to $\Sigma^>$ for $p_0<\omega_F$ and to
$\Sigma^<$ for $p_0>\omega_F$ at zero temperature.}

\end{figure}

The collisional self-energies $\Sigma^\gtrless$ can be calculated
diagrammatically. We restrict ourselves to the lowest order, given by the
direct and the exchange Born diagram (see Fig. \ref{fig:selfenergy}). In this
approximation only two particle correlations are included in the model. The
corresponding processes at finite chemical potential (cf. Ref. \cite{kb}) are shown
in Fig. \ref{fig:scatter}.

In perturbation theory the particle and antiparticle lines in the diagrams of
Fig. \ref{fig:selfenergy} are interpreted as free propagators. Our model,
however, takes into account the in-medium character of the intermediate states.
The lines are interpreted as full relativistic Green's functions which again
depend on the self-energies via the spectral function
(eqs. (\ref{eq:gkl-spec}) and (\ref{eq:ggr-spec})). This way we end up with
self-consistent expressions for the self-energies. In the language of Feynman
diagrams the use of full Green's functions means that we sum over a whole class
of diagrams. Note that no ambiguities concerning Fierz rearrangements appear
here. {\em All} interactions at the two loop level\footnote{Note that using
full propagators instead of free ones implies that only two-particle 
irreducible diagrams have to be considered.} caused by the interaction
term in eq. (\ref{eq:njl-lagr}) are included in our approach. In particular, we find
contributions from quark-quark as well as from quark-antiquark scattering
(cf. Fig. \ref{fig:scatter}).

\begin{figure}[htbp]
	\begin{center}
		\includegraphics{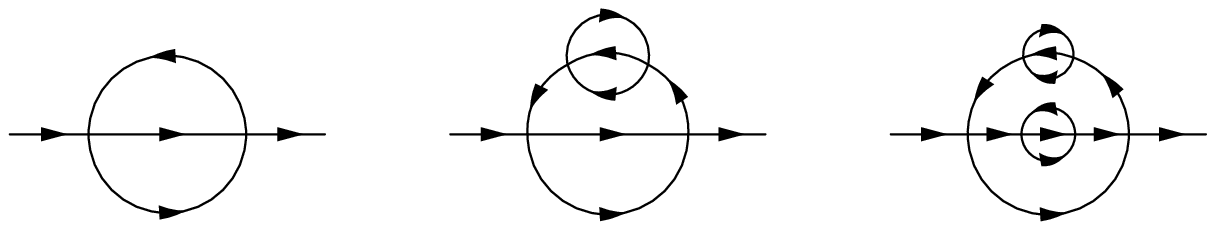}
	\end{center}
	\caption{\label{fig:examples}Examples for diagrams that are resummed in our
	self-consistent calculation. In this figure the lines correspond to free
	propagators. Once the elementary sunset diagram on the left is constructed
	from full propagators the other diagrams are automatically resummed.}
\end{figure}

\begin{figure}[htbp]
	\begin{center}
		\includegraphics[scale=.95]{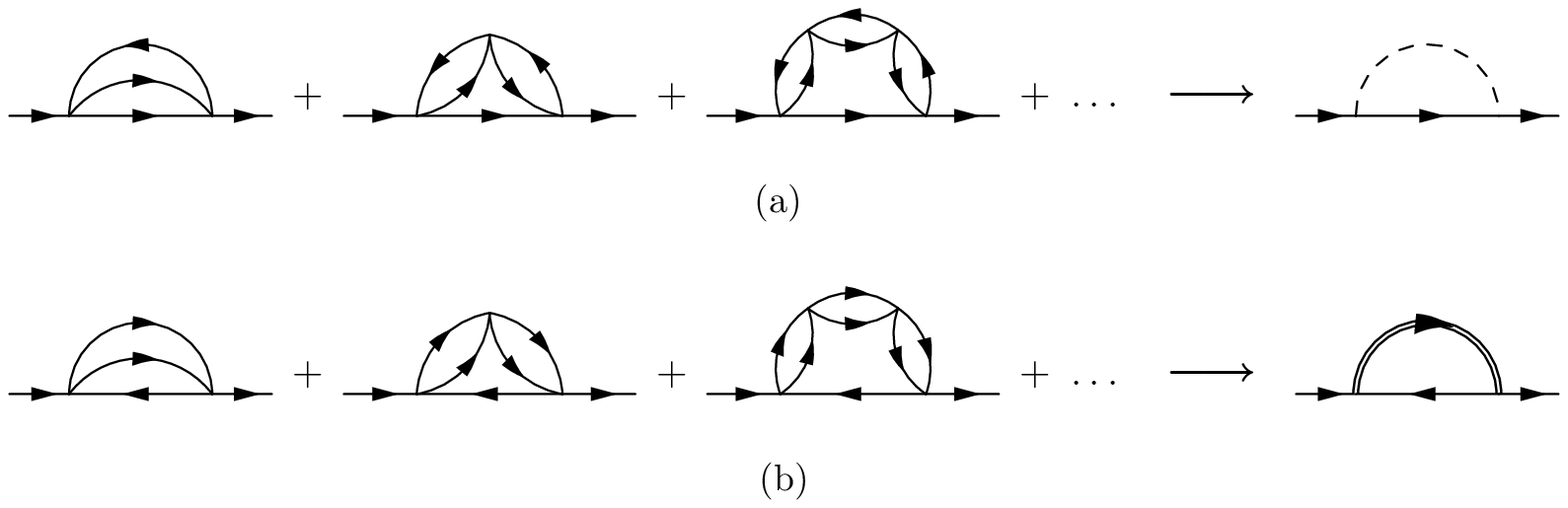}
	\end{center}
	\caption{\label{fig:resummation} Series of diagrams that correspond to the
	coupling of dynamically generated mesons (a) and diquarks (b) to the quarks
	(see \cite{klev,klvw} for details). The dashed line is a pion or sigma and
	the dotted line is a diquark. In the present model only the lowest order
	diagram, the sunset diagram, is included.}
\end{figure}

At this point it might be illuminating to discuss which Feynman diagrams are
resummed by our approach and which ones are not. We use full propagators for
the calculation of the sunset diagram depicted in Fig. \ref{fig:selfenergy}.
In this way we include diagrams like the ones shown in Fig. \ref{fig:examples}.
We do {\em not} sum up the series shown in Fig. \ref{fig:resummation} (a)
(we only have the first term). Series of that type would correspond to the
coupling of dynamically generated pions or sigmas to the quarks
\cite{klev,klvw}. Clearly such series are relevant in the chirally broken
phase. We also do not sum up the series shown in Fig. \ref{fig:resummation}
(b), which corresponds to coupling to diquarks. This is expected to be relevant
if the phenomenon of color superconductivity is tackled.

Working out the diagrams and replacing the Green's functions $g^\gtrless$ by
spectral functions according to eqs. (\ref{eq:gkl-spec}) and
(\ref{eq:ggr-spec}) one gets for the two components of $i\Sigma^>(p)$ in the
chirally restored phase
\bea
	i\Sigma^>_0(p)&=&\frac{G^2}{(2\pi)^6}\int d k_0 dq_0 \int_{<\Lambda}
	 du^2 dv^2  \int d \cos\vartheta_{uv}
	[1-n_F(k_0)] u v \nnl
	&&
	 \Bigg\{ 14 \left[
		\left(\rho_0(k)\rho_0(q)-\rho_\mathrm{v}(q)\rho_\mathrm{v}(k)
		\frac{u^2-v^2}{|\vec u+\vec v||\vec u-\vec v|}
		\right)
		n_F(q_0)[1-n_F(r_0)]
	\right.
	\nnl
	&&\left. \times
		\frac{1}{2|\vec p|v}
		\int^{\vec p^{\,2}+v^2+2|\vec p|v}_{\vec p^{\,2}+v^2-2|\vec p|v}
		dr^2 \rho_0(r_0,r^2)\right]
						_{{\vec k=\frac{1}{2}(\vec u-\vec v)\atop
		\vec q=\frac{1}{2}(\vec u+\vec v)}
		\atop r_0=p_0+q_0-k_0}
		\nnl
	&&- \left[
		\left(\rho_0(k)\rho_0(q)-\rho_\mathrm{v}(q)\rho_\mathrm{v}(k)
		\frac{u^2-v^2}{|\vec u+\vec v||\vec u-\vec v|}\right)[1-n_F(q_0)]n_F(r_0)
	\right.
	\nnl
	&&\left. \times
		\frac{1}{2|\vec p|u}
		\int^{\vec p^{\,2}+u^2+2|\vec p|u}_{\vec p^{\,2}+u^2-2|\vec p|u}
		dr^2 \rho_0(r_0,r^2)\right]
				_{{\vec k=\frac{1}{2}(\vec u-\vec v)\atop
		\vec q=\frac{1}{2}(\vec u+\vec v)}
		\atop r_0=q_0+k_0-p_0}\Bigg\} \label{eq:sigma0}
\eea
and
\bea
	i\Sigma^>_\mathrm{v}(p)&=&\frac{G^2}{(2\pi)^6}\int d k_0 dq_0 \int_{<\Lambda}
	 du^2 dv^2  \int d \cos\vartheta_{uv}
	[1-n_F(k_0)] u v \label{eq:sigmav} \\
		&&
		\Bigg\{ 14 \left[
		\left(\rho_0(k)\rho_0(q)-\rho_\mathrm{v}(q)\rho_\mathrm{v}(k)
		\frac{u^2-v^2}{|\vec u+\vec v||\vec u-\vec v|}\right)n_F(q_0)[1-n_F(r_0)]
	\right.
	\nnl
	&&\left. \times
		\frac{1}{4|\vec p|v}
		\int^{\vec p^{\,2}+v^2+2|\vec p|v}_{\vec p^{\,2}+v^2-2|\vec p|v}
		dr^2 \rho_\mathrm{v}(r_0,r^2)
		\frac{1}{r}\left(\frac{r^2-v^2}{\vec p^2}+1\right)\right]
				_{{\vec k=\frac{1}{2}(\vec u-\vec v)\atop
		\vec q=\frac{1}{2}(\vec u+\vec v)}
		\atop r_0=p_0+q_0-k_0}
		\nnl
	&&+ \left[
		\left(\rho_0(k)\rho_0(q)-\rho_\mathrm{v}(q)\rho_\mathrm{v}(k)
		\frac{u^2-v^2}{|\vec u+\vec v||\vec u-\vec v|}\right)[1-n_F(q_0)]n_F(r_0)
	\right.
	\nnl
	&&\left. \times
		\frac{1}{4|\vec p|u}
		\int^{\vec p^{\,2}+u^2+2|\vec p|u}_{\vec p^{\,2}+u^2-2|\vec p|u}
		dr^2 \rho_\mathrm{v}(r_0,r^2)
		\frac{1}{r}\left(\frac{r^2-u^2}{\vec p^2}+1\right)\right]
		_{{\vec k=\frac{1}{2}(\vec u-\vec v)\atop
		\vec q=\frac{1}{2}(\vec u+\vec v)}
		\atop r_0=q_0+k_0-p_0}\Bigg\}. \nn
\eea
Here $\vec u=\vec q +\vec k$ and $\vec v=\vec q -\vec k$ are the center of mass
and relative momenta in the scattering processes of Fig. \ref{fig:scatter}.
$\vartheta_{uv}$ is the angle between the two vectors $\vec u$ and $\vec v$.
The subscript "$<\Lambda$" at the integrals indicates that a three-momentum
cutoff is applied. This notation requires some comments: The integrals in
eqs. (\ref{eq:sigma0}) and (\ref{eq:sigmav}) are integrals for $\vec u$ and
$\vec v$. In practice, however, the cutoff is applied to $\vec k$ and $\vec q$
in the calculations. The vectors $\vec k$ and $\vec q$ are the momenta of the
particles and antiparticles in Fig. \ref{fig:selfenergy}, thus they must be
regularized when working out the diagrams.

The expressions for $-i\Sigma^<_{0}(p)$ and $-i\Sigma^<_{\mathrm{v}}(p)$ are
not presented here. They are easily found from eqs. (\ref{eq:sigma0}) and
(\ref{eq:sigmav}) by replacing all functions $n_F$ by $(1-n_F)$ and vice versa.

Eqs. (\ref{eq:sigma0}) and (\ref{eq:sigmav}) relate the collisional
self-energies (collision rates) to the spectral function. Note that the zeroth
and the vector component of eq. (\ref{eq:spec-struc}) are mixed. Finally, the
components of the spectral function have to be expressed in terms of the
self-energies $\Sigma^\gtrless$ via the width $\Gamma$. The expressions are
found by inserting the explicit form of the relativistic in-medium Green's
function,
\bea
	g^\mathrm{ret}(p)=\frac{1}{\not\!p-\Sigma^\mathrm{ret}(p)},
	\label{eq:ret-prop}
\eea
into the definition of the spectral function (\ref{eq:spec-def}). Since the
chirally restored phase is considered the quarks are massless and no mass term
appears in the denominator.

When working out the imaginary part of $g^\mathrm{ret}$ one also needs the real
part of the self-energy $\Sigma^\mathrm{ret}$. In principle it can be
calculated from $\Gamma$ by use of the dispersion relation (\ref{eq:disp-rel}).
The studies of Lehr et al. \cite{le} have shown that the influence of
$\mathrm{Re}\Sigma^\mathrm{ret}$ on the nucleon spectral function and the
nucleon momentum distribution in  nuclear matter is small. In their first
calculations they have replaced $\mathrm{Re}\Sigma^\mathrm{ret}(p)$ by a
constant value independent of energy and momentum. It was found that this
violation of analyticity leads only to minor effects in the results.

In our present calculations, we are not able to use the dispersion relation
(\ref{eq:disp-rel}) due to technical reasons. We work on a finite grid in
energy and momentum space, the boundaries of which have been chosen to include
all significant parts of the spectral function. The width $\Gamma_0$, however,
extends much further into the high energy regions than the spectral function.
As we do not know $\Gamma_0$ outside the grid, the $p_0$ integration of the
dispersion integral cannot be performed\footnote{Keep in mind that the NJL
cutoff is a three-momentum cutoff. While all functions are zero at momenta
above the cutoff there is no direct consequence for high energies. In fact the
cutoff does affect the high energy behavior of $\Gamma_0$
(cf. eqs. (\ref{eq:sigma0}) and (\ref{eq:sigmav})) but only in the form of a
contiuous decline (we will come back to that point at the discussion of the
results).}.

In principle, one could find $\mathrm{Re}\Sigma^\mathrm{ret}$ in a different
way. Using the dispersion relation for the retarded propagator,
$\mathrm{Re}g^\mathrm{ret}$ can be calculated from the spectral function.
Then it is possible to deduce $\mathrm{Re}\Sigma^\mathrm{ret}$ from
eq. (\ref{eq:ret-prop}). We have checked this and it turned out that the real
part of the self-energy is very small compared to $p_0$ and $|\vec p|$, i.e.,
in $\not\!p-\mathrm{Re}\Sigma^\mathrm{ret}$ the dominant part is $\not\!p$.
Since we know from the nucleons that the real part of $\Sigma^\mathrm{ret}$ is
relatively unimportant $\mathrm{Re}\Sigma^\mathrm{ret}$ is set to zero in the
following.

The following expressions for $\rho_0$ and $\rho_\mathrm{v}$ are found from
eqs. (\ref{eq:spec-def}) and (\ref{eq:ret-prop}), when
$\mathrm{Re}\Sigma^\mathrm{ret}$ is set to zero
\bea
		\rho_0(p_0,\vec p^{\,2})
		&=&  \frac{(p_0^2+\vec p^{\,2})\Gamma_0
		+\frac{1}{4}(\Gamma_0^2-\Gamma_\mathrm{v}^2)\Gamma_0
		+2p_0|\vec p|\Gamma_\mathrm{v}}
		{[p_0^2-\vec p^{\,2}
		-\frac{1}{4}\Gamma_0^2+\frac{1}{4}\Gamma_\mathrm{v}^2]^2
		+[p_0\Gamma_0-\Gamma_\mathrm{v}|\vec p|]^2}, \label{eq:rho0-gamma}
\\
		\rho_\mathrm{v}(p_0,\vec p^{\,2})
		&=& - \frac{(p_0^2+\vec p^2)\Gamma_\mathrm{v}-\frac{1}{4}
		(\Gamma_0^2-\Gamma_\mathrm{v}^2)\Gamma_\mathrm{v}
		+2p_0|\vec p|\Gamma_0}
		{[p_0^2-\vec p^{\,2}
		-\frac{1}{4}\Gamma_0^2+\frac{1}{4}\Gamma_\mathrm{v}^2]^2
		+[p_0\Gamma_0-\Gamma_\mathrm{v}|\vec p|]^2}. \label{eq:rhov-gamma}
\eea
The two relations look very similar. Up to some signs only $\Gamma_0$ and
$\Gamma_\mathrm{v}$ are exchanged in the numerator. Near the on-shell point
$\rho_0$ and $\rho_\mathrm{v}$ differ only by a minus sign\footnote{Concerning
the sign of $\rho_\mathrm{v}$ recall our remark after eq.
(\ref{eq:spec-struc}).} at positive $p_0$ if terms $\sim \mathcal{O}(\Gamma^3)$
are neglected
\bea
		\rho_0(p_0,p_0^2)
		\approx \frac{2 p_0^2\Gamma_0
		+2p_0|p_0|\Gamma_\mathrm{v}}
		{\cdots} ~,
		\qquad
		\rho_\mathrm{v}(p_0,p_0^2)
		\approx - \frac{2 p_0^2\Gamma_\mathrm{v}
		+2p_0|p_0|\Gamma_0}
		{\cdots}. \label{eq:rho-approx}
\eea
Eqs. (\ref{eq:sigma0}) and (\ref{eq:sigmav}), (\ref{eq:rho0-gamma}) and
(\ref{eq:rhov-gamma}), and eq. (\ref{eq:gamma-sigma}) form a set of equations
describing a self-consistency problem. A direct solution of this problem is not
possible but the equations can be used to calculate the spectral function
iteratively.

\section{Numerics and results\label{sec:results}}

\subsection{Details of the calculation}

All calculations were performed at zero temperature in the chirally restored
phase. We have used two different quark densities. Motivated from the
investigations for nucleons in nuclear matter, we chose the quark matter in the
first case such that it is comparable to regular nuclear matter. As every
nucleon consists of three valence quarks the quark density was set to
$\rho_\mathrm{qm}=3\cdot\rho_\mathrm{nm}=3\cdot 0.17 \,\mathrm{fm}^{-3}$. For
the number of flavors we use $N_\mathrm{f}=2$ (up and down). This yields a
Fermi energy and a Fermi momentum of $\omega_F=p_F=0.268\,\mathrm{GeV}$ for
massless quarks. In reality quark matter with that chemical potential would not
be in the chirally restored phase. Therefore, we have also worked with a three
times higher density, $\rho_\mathrm{qm}=1.53 \,\mathrm{fm}^{-3}$, corresponding
to a Fermi energy of $\omega_F=0.387\,\mathrm{GeV}$. This case seems to be more
realistic for chirally symmetric quark matter, the chemical potential is well
beyond the chiral phase transition in the NJL model \cite{klev,klvw}. Note that
all states at energies up to $\omega_F$ are filled. This includes the states at
negative energies. Thus there are no holes in the Dirac sea which could be
identified with populated antiquark states.

The three-momentum cutoff $\Lambda=653\,\mathrm{MeV}$ and the coupling constant
$G=2.14\cdot\Lambda^{-2}$ of the NJL model were chosen so that the model gives
the known values for the quark condensate,
$\langle\bar u u\rangle=\langle\bar d d\rangle=-(250\,\mathrm{MeV})^{3}$, and
the pion coupling constant, $f_\pi =93\,\mathrm{MeV}$, in vacuum \cite{klev}.
In the case of the lower density we performed also calculations with two times
and four times larger coupling strengths to investigate the influence on the
spectral function.

The calculations were carried out on an energy-momentum grid with boundaries
$-653\,\mathrm{MeV}\leq p_0\leq 653\,\mathrm{MeV}$ and
$|\vec p|\leq 653\,\mathrm{MeV}$ (in correspondence to the NJL cutoff) and a
mesh size of $6.53\,\mathrm{MeV}$ in both directions. Due to the cutoff a grid
of that size is sufficient to include all significant parts of the spectral
function. The states with negative energies $p_0$ are interpreted as anti-quark
states with positive energies $|p_0|$ in the discussion of the results.

To initialize the calculations  a constant width, $\Gamma_0=1\,\mathrm{MeV}$
and $\Gamma_\mathrm{v}=0$, was used. In principle it would be interesting to
start from a quasiparticle approximation with a width much smaller than the
final result. This would allow to observe the redistribution of strength away
from the peaks during the iterations. However, this is numerically not
feasible. So the initialization values are not too far away from the final
results and only two iterations were necessary to reach self-consistency.

\subsection{Results}

In this section, we present the results of our iterative calculations. Before
the spectral function is discussed we have a look at the width $\Gamma$. This
is helpful to understand the resulting spectral functions and it allows a
clearer observation of the iterative process than the spectral function itself.

We show the results for $\rho_0$ and $\Gamma_0$ but not for $\rho_\mathrm{v}$
and $\Gamma_\mathrm{v}$. Most of the interesting structure of the spectral
function is located around the on-shell peaks. It has been discussed that
$\rho_0$ and $\rho_\mathrm{v}$ differ at most by a sign near those peaks
(eq. (\ref{eq:rho-approx})). Thus it is not surprising that both look very
similar and it is sufficient to show only one of them. The reason to show
$\rho_0$ is that the density of states (particles minus antiparticles) is
given by $\rho_0$ alone,
\bea
\mathrm{Tr}\{\gamma_0 \mathcal{A}(p)\}= 4\cdot N_\mathrm{f} N_\mathrm{c}
\rho_0(p_0,\vec p^{\,2}).
\eea
There is no simple relation between $\Gamma_0$ and $\Gamma_\mathrm{v}$
justifying to show only one. But our results show that $\Gamma_\mathrm{v}$
is much smaller than $\Gamma_0$ (this has also been observed in calculations
for relativistic nucleons \cite{horo}). Hence $\Gamma_\mathrm{v}$ seems rather
uninteresting since its influence on the spectral function is minimal.

\begin{figure}[htbp]
	\begin{center}
		\includegraphics[scale=.72]{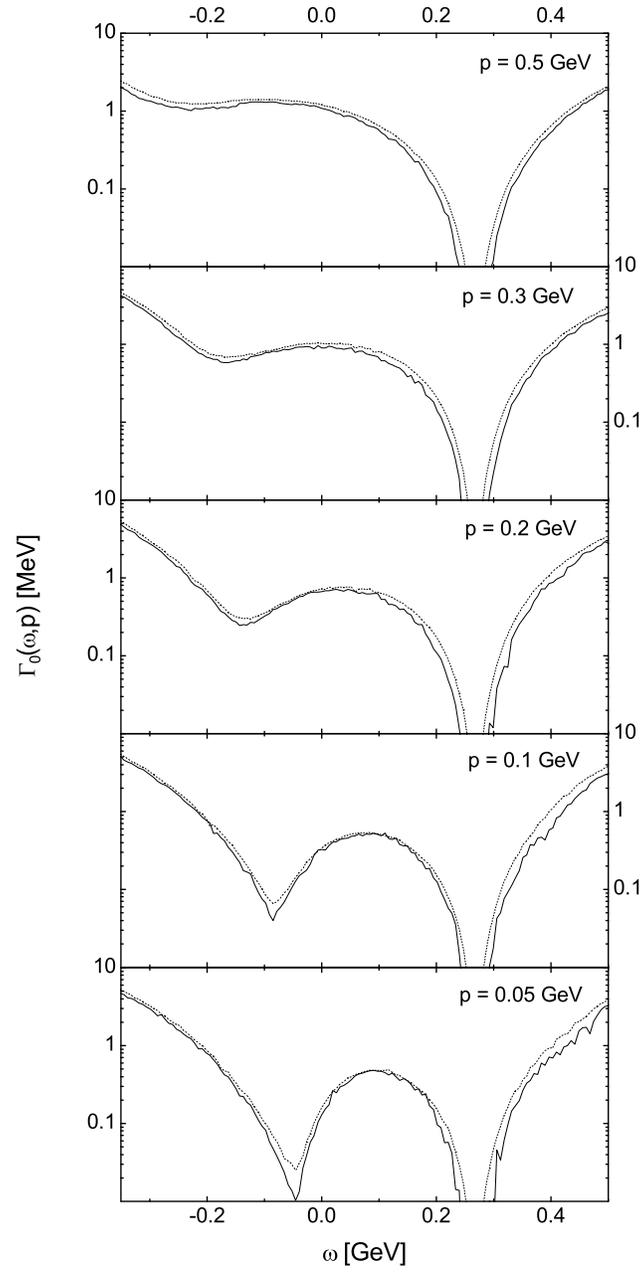}
	\end{center}
	\caption{\label{fig:g0series}Width of the spectral function of quarks at
	different momenta. The dashed line shows the result after the first
	iteration, the solid line shows the final result after two iterations.}
\end{figure}

Figure \ref{fig:g0series} shows cuts of the width $\Gamma_0$ at several momenta
for $\omega_F=0.268\,\mathrm{GeV}$. The dotted line is the result after the
first iteration while the solid line displays the final, self-consistent
result. Physically the most interesting area lies in the energy range
$0<p_0<\omega_F$ since that is the location of the populated quark states. All
states above the Fermi energy as well as the antiquark states at negative
$p_0$ are unoccupied (no holes in the Dirac sea). $\Gamma_0$ takes on low
values of $0.1-1\,\mathrm{MeV}$ in this zone, being close to the initialization
value.

The two most important features of $\Gamma_0$ are clearly visible. First, the
width is zero along the Fermi energy $\omega_F$, the quarks along this line are
quasiparticles. Due to Pauli blocking and energy conservation it is not
possible to scatter into or out of this configurations at zero temperature.
Second, the width grows explosively at high $|p_0|$. The reason for this is
the pointlike interaction of the NJL model. The self-energies $\Sigma^\gtrless$
essentially sum up the phase space available for scattering processes. The
opening of this phase space at large $|p_0|$ can be directly read off from
$\Gamma_0$. The three-momentum cutoff applied in the calculations stops this
inflational behavior at higher energies. These energies, however, are outside
of our grid. A simple (on-shell) estimate shows that the maximum width is
reached at $\Lambda<|p_0|<2\Lambda$ and that the width must be zero for
$|p_0|>3\Lambda$ .

Another noticeable feature is the local minimum in the antiparticle sector,
most visible at low momenta. This is a phase space effect that can be easily
explained by the energy and momentum dependence of the particle correlations.
In the bottom row of Fig. \ref{fig:scatter} all processes contributing to the
width at negative $p_0$ are shown. Phase space opens for the processes
$\bar p k\rightarrow \bar q r$ and $\bar p\rightarrow \bar k\bar q r$ at
$p_0=0$ and $p_0=-\omega_F$, respectively, and grows with decreasing $p_0$. On the other
hand the process $\bar p k q \rightarrow r$ is possible for $p_0>-|\vec p|$
only. Its phase space grows with increasing $p_0$ and is maximal at $p_0=0$.
The overlap of these three processes leads to the observed minimum that is
situated approximatly at $p_0=-|\vec p|$ for low momenta.

\begin{figure}[htbp]
	\begin{center}
		\includegraphics[scale=.7]{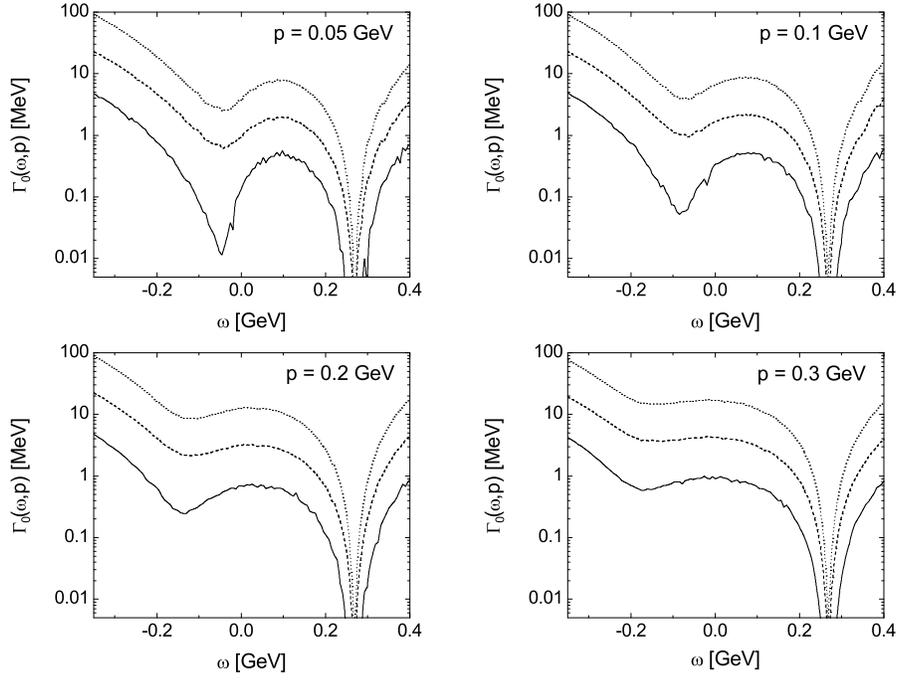}
	\end{center}
	\caption{\label{fig:g0coupl}Width of the spectral function of quarks at
	different momenta. The solid line shows the width at the usual coupling
	strength ($G=2.14\cdot\Lambda^{-2}$), the dashed line is the result at a
	coupling twice as large and the dotted line has been obtained with a
	coupling four times larger than the usual value.}
\end{figure}

In Fig. \ref{fig:g0coupl}, the width $\Gamma_0$ is shown for
$\omega_F=0.268\,\mathrm{GeV}$ at several values for the coupling strength;
all curves are self-consistent final results. The solid lines show the result
at the original coupling of $G=2.14\cdot\Lambda^{-2}$, the dashed lines were
obtained for $G=2\cdot2.14\cdot\Lambda^{-2}$ and the dotted lines display the
width at a coupling strength of $G=4\cdot2.14\cdot\Lambda^{-2}$. The larger
couplings lead to a significant increase of the width. Since $G$ enters the
self-energies as a factor of $G^2$ it is not surprising that a two times larger
coupling results in a four times larger width and a four times larger coupling
causes a width 16 times larger. The shape of $\Gamma_0$ does not change very
much. Only the gap in the antiquark sector gets smeared out due to the
increased width which weakens the restrictions for the scattering processes.

\begin{figure}[htbp]
	\begin{center}
		\includegraphics[scale=.7]{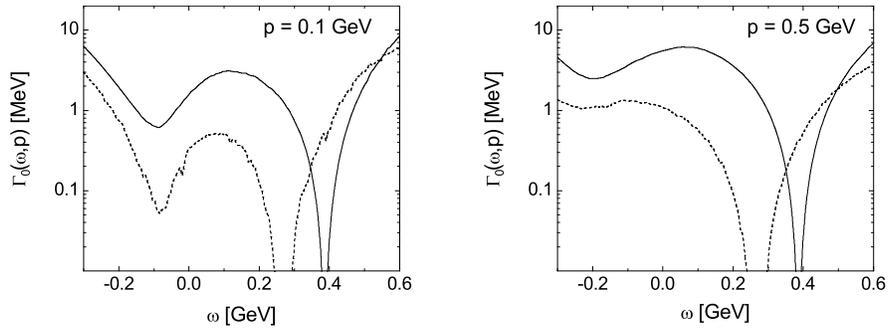}
	\end{center}
	\caption{\label{fig:g0mu}Width of the spectral function at two different
	momenta. The solid line shows the result at a chemical potential of
	$\omega_F=0.387\,\mathrm{GeV}$ while the dashed line has been found for
	$\omega_F=0.268\,\mathrm{GeV}$ (cf. Fig. \ref{fig:g0coupl}). In both cases a
	coupling of $G=2.14\cdot\Lambda^{-2}$ has been used.}
\end{figure}

Figure \ref{fig:g0mu} displays $\Gamma_0$ for the different chemical potentials
$\omega_F=0.268\,\mathrm{GeV}$ and $\omega_F=0.387\,\mathrm{GeV}$ at the
regular coupling strength. The higher chemical potential has a similar effect
as an increased coupling. Essentially, a scaling of the width can be observed
while the general shape of $\Gamma_0$ remains unaffected. For $p_0>\omega_F$
one would expect a width $\Gamma_0\sim \rho$. This is, however, hard to see
because of the vicinity to the cutoff in our calculations. For $p_0<\omega_F$,
i.e., for the hole width, one expects a width $\Gamma_0\sim\rho^2$
(see Fig. \ref{fig:scatter}, second line), because both momenta $k$ and $q$
lie within the Fermi sea. The observed scaling, however, is slightly smaller.
This is due to the fact that the cutoff suppresses outgoing states with
energies $r_0>\Lambda$ that can be reached when $\omega_F>\Lambda/2$. So the
widths in our results differ not by a factor of $9$ but only by a factor of
$\sim 6$ as one would expect for a Fermi energy of $\omega_F=\Lambda/2$ (this
is the largest Fermi energy possible for which no outgoing states are
suppressed).

\begin{figure}[htbp]
	\begin{center}
		\includegraphics[scale=.7]{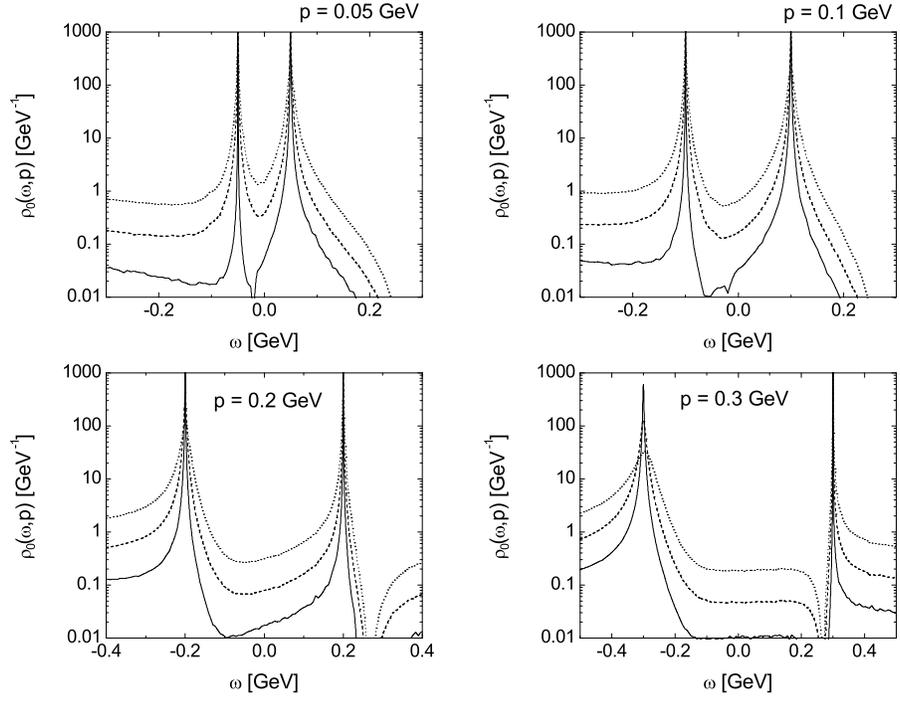}
	\end{center}
	\caption{\label{fig:rho0coup}The spectral function of quarks at different
	momenta and couplings (cf. Fig. \ref{fig:g0coupl} for details).}
\end{figure}

The spectral function defines the spectrum of possible energies $p_0$ for a
particle or antiparticle with momentum $\vec p$ that is added to the medium.
In Fig. \ref{fig:rho0coup}, $\rho_0$ is shown for the three coupling strengths
at the lower chemical potential of $\omega_F=0.268\,\mathrm{GeV}$.  The
structure is clearly dominated by the two peaks at $p_0=|\vec p|$ and
$p_0=-|\vec p|$. These are the on-shell peaks of the quarks and the
antiquarks. Because of the small widths they are very narrow and most of the
strength is concentrated there. As one would expect from the results for
$\Gamma_0$ the peaks get broader and less high when the coupling is increased.
Thus strength is removed from the peaks to the off-shell regions. The
background of the spectral function seems to scale with the coupling in a
similar way as $\Gamma_0$. Approximately a quadratic dependence is found.

The spectral function for the quarks and the antiquarks is not symmetric. This
is due to the finite chemical potential: The quarks of our medium fill up all
states up to the Fermi energy.  Due to Pauli blocking only scattering processes
with outgoing quarks at energies above $\omega_F$ are allowed. Incoming quarks
from the medium must always have energies below $\omega_F$. On the other hand,
all antiquark states are empty. Scattering processes with outgoing antiquarks
at all energies are possible. However, there are no processes with incoming
antiquarks from the medium.

Since we are interested in short-range correlations the spectral function at
large momenta above the Fermi momentum is of particular interest. The cut at
$|\vec p|=0.3\,\mathrm{GeV}$ in Fig. \ref{fig:rho0coup} indicates that
short-range correlations seem rather weak for the original coupling strength,
the spectral function is very small at $0<p_0<\omega_F$. Only when the coupling
is increased a significant population of high momentum states can be observed.

\begin{figure}[htbp]
	\begin{center}
		\includegraphics[scale=.7]{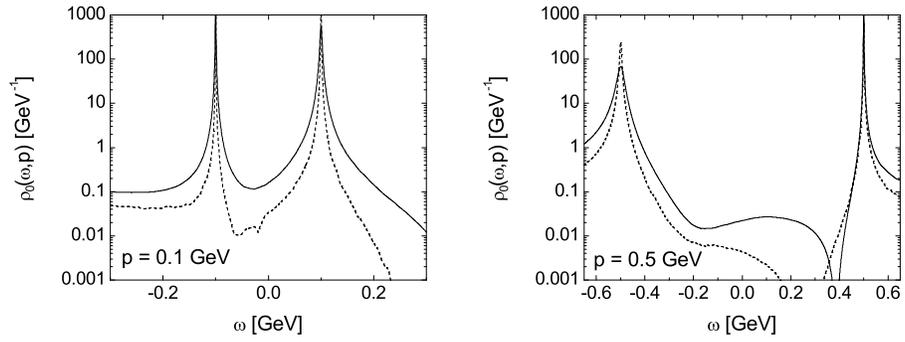}
	\end{center}
	\caption{\label{fig:rho0mu}The spectral function for two different chemical
	potentials and momenta (see Fig. \ref{fig:g0mu} for details).}
\end{figure}

In Fig. \ref{fig:rho0mu}, the influence of the chemical potential on the
spectral function is illustrated. As expected from the results for the width
the peaks get broader and less high for the larger chemical potential of
$\omega_F=0.387\,\mathrm{GeV}$. Of particular interest is the cut for
$|\vec p|=0.5\,\mathrm{GeV}$ (i.e., $p>p_F$). A bump in the region of the
occupied quark states indicates the growing importance of short-range
correlations for higher $\omega_F$ (one should compare this bump to the cuts
for $|\vec p|=0.3\,\mathrm{GeV}$ in Fig. \ref{fig:rho0coup} to take into
account the different Fermi momenta).

\begin{figure}[htbp]
	\begin{center}
		\includegraphics[scale=.4]{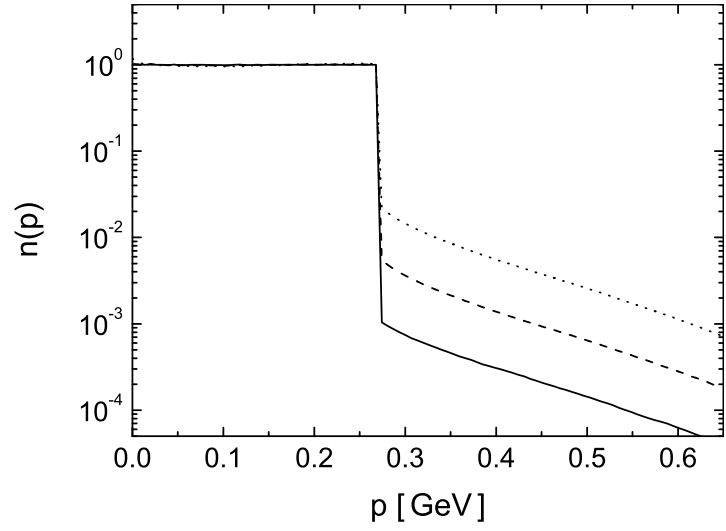}
	\end{center}
	\caption{\label{fig:momdist}Quark momentum distribution in quark matter at
	different coupling strengths (see Fig. \ref{fig:g0coupl} for details).}
\end{figure}

\begin{figure}[htbp]
	\begin{center}
		\includegraphics[scale=.4]{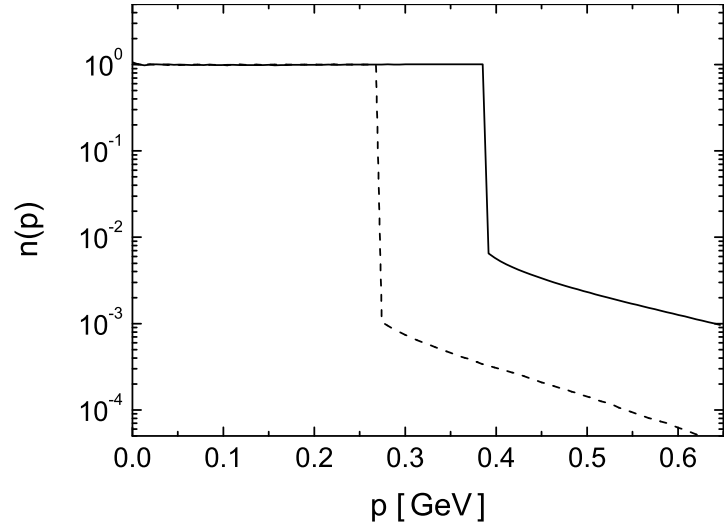}
	\end{center}
	\caption{\label{fig:momdistmu}Quark momentum distribution in quark matter
	for two different chemical potentials (see Fig. \ref{fig:g0mu} for details).}
\end{figure}

Finally we take a look at the (normalized) momentum distribution of the quarks
which is given by
\bea
	n(|\vec p|)=\int_0^{\omega_F}\frac{dp_0}{\pi} \rho_0(p_0,|\vec p|).
\eea
The momentum distribution of nucleons in nuclear matter or nuclei shows a
depletion of occupation probabilites by about 10\% \cite{le}. The resulting
high-momentum tail is taken as a universal sign of short-range correlations.
Figure \ref{fig:momdist} displays the quark momentum distribution at the
different coupling strengths for $\omega_F=0.268\,\mathrm{GeV}$. As expected
for an infinite system a sharp step at the Fermi momentum appears. At the
lowest coupling a depletion of only 0.1\% is found. This confirms the previous
interpretation of the spectral function. For the coupling twice as large the
short-range correlations increase but still the depletion effect is below 1\%.
Only for the largest couplings we find a high-momentum tail of a few percent,
comparable to the case of nucleons. Figure \ref{fig:momdistmu} shows again that
the larger chemical potential has an effect similar to the increased couplings.
The depletion for $\omega_F=0.387\,\mathrm{GeV}$ grows by almost one order of
magnitude compared to the case of $\omega_F=0.268\,\mathrm{GeV}$ and the
original coupling strength. It has a size of almost 1\% and is comparable to
the result for the two times larger coupling.

\section{\label{sec:summary}Summary and conclusions}

In this paper we have presented an approach to the spectral function of quarks
in quark matter. Based on a successful concept for nuclear matter it was shown
how to construct a simple model for the quarks. Using the relations between
collisional self-energies and the spectral function an iterative method was
derived that goes beyond the quasiparticle approximation. Assuming that the
exact structure of the interaction is irrelevant for the calculations -- an
important finding for nuclear matter --  the pointlike interaction of the NJL
model was used in the calculations.

Our calculations, which are currently restricted to zero temperature and the
chirally restored phase, show that we are able to deal with the numerical
difficulties implied by this approach. However, the results also indicate that
the influence of short-range correlations is small compared to nuclear matter.
This finding, however, might be an artefact of the present model, the NJL model
with vacuum parameters in the Born approximation. Since we know now that the
model is technically feasible we can go beyond the simple model presented here.
It is our plan to use more sophisticated interaction models for future
calculations and to explore also other phases with broken symmetries.


\vspace{4cm}


\end{document}